\pdfoutput=1
\documentclass[aps,pre,10pt,twocolumn,superscriptaddress,balancelastpage,showpac
s,floatfix]{revtex4-1}

\usepackage[pdftex]{graphicx}

\usepackage[centertags]{amsmath}
\usepackage{amsfonts}
\usepackage{version}
\usepackage{euscript}
\usepackage{amssymb}
\usepackage{amsthm}
\usepackage{newlfont}
\usepackage{stmaryrd}
\usepackage{mathrsfs}
\usepackage{subfigure}
\usepackage{hyperref}
\usepackage{color}

\newcommand{\id}{\textrm{d}}
\newcommand{\vdot}{\dot{v}}
\newcommand{\eq}{Eq.}
\newcommand{\eqs}{Eqs.}
\newcommand{\cP}{{\cal P}}
\newcommand{\cS}{{\cal S}}
\newcommand{\cT}{{\cal T}}
\newcommand{\cA}{{\cal A}}
\newcommand{\cO}{{\cal O}}
\newcommand{\tD}{{\widetilde D}}
\newcommand{\tC}{{\widetilde C}}

\providecommand{\avg}[1]{\left \langle #1 \right \rangle}
\providecommand{\pnt}[1]{\left ( #1 \right)}

\newcommand{\executeiffilenewer}[3]{%
  \ifnum\pdfstrcmp{\pdffilemoddate{#1}}%
  {\pdffilemoddate{#2}}>0%
  {\immediate\write18{#3}}\fi
}

\begin{document}
\title{Fluctuation-response relations for nonequilibrium diffusions
  with memory}

\author{C. Maes}
\author{S. Safaverdi}
\affiliation{Instituut voor Theoretische Fysica, KU Leuven - Celestijnenlaan 200D, B-3001 Leuven, Belgium}

\author{P. Visco}
\affiliation{Laboratoire Mati\`ere et Syst\`emes Complexes, CNRS UMR
  7057, Universit\'e Paris Diderot, 10 rue Alice Domon et L\'eonie
  Duquet, 75205 Paris cedex 13, France}
\author{F. van Wijland}
\affiliation{Laboratoire Mati\`ere et Syst\`emes Complexes, CNRS UMR
  7057, Universit\'e Paris Diderot, 10 rue Alice Domon et L\'eonie
  Duquet, 75205 Paris cedex 13, France}

\date{\today}

\begin{abstract}
Strong interaction with other particles or feedback from the medium on a Brownian particle entail memory effects in the effective dynamics.
That motivates the extension of the fluctuation-dissipation theorem to
nonequilibrium Langevin systems with memory.   An important application is to the nonequilibrium modification of the Sutherland-Einstein relation between diffusion
 and mobility in the case of strong memory. Nonequilibrium corrections include the time-correlation between the dynamical activity and the velocity of the particle, which in turn leads to information about the correlations 
between the driving force and the particle's displacement.
\end{abstract}

\maketitle

\section{Introduction}

Path-integrals are robust against small perturbations in the dynamics
and hence, make expansions easier such as for the derivation of
response relations.  That has been systematically applied before for
the extension of the fluctuation-dissipation theorem to nonequilibrium
systems, at least in the Markov case \cite{har,up,proc,fdr}.  The
purpose of the present paper is to further extend the nonequilibrium
linear response theory to dynamics with memory, which is physically
often more appropriate, and which has been so far considered only
in a ``weak'' non-Markovian case, where memory decreases exponentially
fast~\cite{villamaina,crisanti}. Also here, the very name
fluctuation-{\it dissipation} relation needs to be revised and perhaps
altered, as the response obtains correlations both with the excess
entropy flux (which is responsible for the standard relation with
dissipation) and with the time-symmetric part of the action, which has
been called the {\it frenetic} contribution as it ultimately relates
to the dynamical activity in the process, \cite{fdr}.  For more
practical purposes it is the latter frenetic contribution where the
steady nonequilibrium forcing appears and hence, fluctuation-response
relations can yield information about that forcing.  Applications such
as to the mobility of particles in living cells are in
progress \cite{pro}, but in the present paper we concentrate on the
general framework, numerical exploration and some technical details.\\

An important ingredient in the present work is the presence of memory
in the equations of motion of a colloidal particle.  The origin of
memory is diverse but it is always related to coupling with other
particles and/or with the environment.  In the case of dense colloidal
suspensions, the reduced dynamics of a single particle certainly
contains memory by integrating out the other particles.  The
theoretical study of the relation between its diffusion and its
mobility is therefore advanced by the analysis of generalized Langevin
equations (GLE).  The latter also appear from other reduction and
projection schemes as generally treated via Zwanzig-Mori
techniques \cite{zwa61, mor65}.  There, temporal scale separation or
micro-macro transfer are the important considerations, but also
intrinsic properties of the medium can contribute memory effects.  For
the latter we have in mind visco-elastic media which react back on
active particles from their previous history.  These are of special
interest for mesoscopic processes in tissues or membranes within
living organisms which are known to respond more easily to external
loads.\\ Here we do not concentrate on the specific interactions or
mechanism that have created the memory effects but we start from
driven GLE for which we assume a structure that is relevant for a
large number of cases of suspensions under the influence of external
forces.  See \cite{pan} for a more recent microscopic--based
derivation of GLE in the context of polymer physics.  The most
important element in our modeling scheme is the principle of local
detailed balance.  It derives from the underlying microreversibility
which gives a strong connection between entropy flux and time-reversal
breaking, \cite{mae03}.  As we will see in the next section,
application of local detailed balance leads to the so called Einstein
relation, also called second fluctuation-dissipation relation, between
friction and noise in the GLE \cite{kub66}, even when modeling
nonequilibrium situations.  We do however not pay special attention to
the choice of driving but we are interested here in general features
and structures of the response.  For a specific application of these
general methods we refer to the recent work \cite{pro} on
reconstructing the active forces from quantitative information on the
violation of the fluctuation-dissipation theorem.  We believe however
that many more applications are waiting, in fact in all these cases
where one can measure deviations from the standard
Kubo-theory \cite{tod92,kub66}.

The study of response in nonequilibrium suspensions is of course not
new, see e.g. \cite{brad,krug1,krug2,ber} for applications to sheared
media. In that respect the present contribution starts from GLE and investigates what are the general
structures that determine the linear response. In particular the work of \cite{camillep, camillet} addresses very similar questions and uses Martin-Siggia-Rose field theory to obtain a fluctuation-response theory.
We concentrate on Sutherland-Einstein relations and we emphasize the structure in terms of entropic versus frenetic contributions, making contact with the Markov formulation in \cite{proc}.
Close to the present work are also the results relating
energy dissipation to the difference of the response and velocity
correlation functions, \cite{har,deu} also for GLE.  Here we emphasize
however the modified Sutherland-Einstein relation connecting diffusion
and mobility.

In the next section the set-up is considered for generalized Langevin
systems with Gaussian noise.  They are driven away from equilibrium by
non-conservative forces. We derive the linear response relations in
Section \ref{linex}.  These are new results, ready to be applied in a
new relation between diffusion and mobility for colloidal particles in
nonequilibrium visco-elastic media.  Section \ref{secse} contains the
simulation results for exploring the modified Sutherland--Einstein
relation and adds visual information on the behavior of the various
terms in the modified relation. The main result of the paper is the
extension of the work in \cite{proc} to include (even strong) memory
effects and to be explicit also about the relevance of the
correlations with dynamical activity and with the forcing.

\section{Theory}\label{theo}

\subsection{Set-up}
Consider the Langevin equation for the position $x_t$ and the velocity
$v_t$ of a (mass 1) particle in a medium at uniform temperature:
\begin{eqnarray}\label{vin}
\frac{\id x_t}{\id t} & = & v_t  \\ \frac{\id v_t}{\id t} & =
& - \int \id s \; \gamma(t-s) v_s + F_t(x_t) +
\sqrt{\frac{2}{\beta}} \,\eta_t + h_t
\nonumber
\end{eqnarray}
To lighten the notation we shall consider that, unless otherwise
specified, integral bounds are understood to range from $-\infty$ to
$\infty$. We take the memory kernel $\gamma(t)\geq 0$ to be causal:
$\gamma(t)=0$ for $t<0$. The Markov case with friction coefficient
$\gamma >0$ is recovered whenever $\gamma(|t|)=2\gamma \,\delta(t)$ is
proportional to the Dirac delta function, which can be achieved for
example from $\gamma(t) = \gamma\,\alpha \exp(-\alpha t) \Theta(t)$ in
the $\alpha \to \infty$ limit, with $\Theta(t)$ the Heaviside step
function. The $F_t$ is the forcing, possibly time-dependent and
non-conservative. It can include effective randomness beyond the
Gaussian noise $\eta_t$, as e.g. in~\cite{pro}.  The parameter
$\beta$ is the inverse temperature of the environment, which we have
taken in front of the noise $\eta_t$. The force $\eta_t$ is a
stationary Gaussian noise-process with zero mean.  We wait to describe
its time-correlations --- see formula \eqref{loc_det_bal} below.  The
last term $h_t = f_t \,\Theta(t)$ is a time-dependent (small)
perturbation --- we will linearly expand around $f_t=0$.  For
simplicity we use a one-dimensional notation, also in what follows,
but the extension to other geometries or dimensions, sometimes
essential for nonequilibrium effects, is straightforward.  We will not
use a Fokker-Planck description in what follows (but we use path
integrals); actually the relation between generalized Langevin and
generalized Fokker-Planck equations in the presence of
position-dependent forces is not entirely clear and to our knowledge
no results have been added after 1980 --- see \cite{han1,fox} for what
we do know.\\

For path-space integration we need some further notation.  Easiest is
to take doubly-infinite paths $\omega = \big(x_s, v_s, \,-\infty < s <
+\infty\big)$.  The price to pay is that some expressions (integrals)
become rather formal.
We refer to \cite{han} for a more detailed reference.  Feasible
alternatives or complements to path-integration to derive
fluctuation--response relations for non-Markovian processes are known
as Furutsu-Novikov theorems, see. e.g \cite{deu,nov,fur,don}.\\
Because the noise $\eta_t$ is a stationary Gaussian process the
path-space measure is completely determined by the symmetric kernel
$\Gamma(t)$ for which
\begin{equation}\label{del}
 \int \id
 s \,\Gamma(t-s) \,\left< \eta_s \eta_r \right> = \delta(t-r)
\end{equation}
The weight of a path $\omega$ is then proportional to
\begin{equation}
\cP_h(\omega)\propto \exp{- \frac{\beta}{4}\int \id
s \int \id r \, \Gamma(r-s)}
\eta_s\,\eta_r
\end{equation}
with
\begin{equation*}
\eta_s = \vdot_s+ \int \id u \,\gamma(s-u) \,v_u - F_s(x_s) -  h_s
\end{equation*}
Compared with the unperturbed dynamics ($h_t\equiv 0$) we have
\begin{equation}\label{act}
\cP_h(\omega) = \cP_0(\omega) \,e^{-\cA_h(\omega)}
\end{equation}
with action
\begin{widetext}
\begin{multline}
 \cA_h(\omega) =- \frac{\beta}{2} \int \id s
 \int \id r \Gamma(r-s) h_s \vdot_r -\frac{\beta}{2}
 \int \id s \int \id r
 \int \id u \,\Gamma(r-s) \,\gamma(r-u)\,v_u\, h_s
\\+ \frac{\beta}{2} \int \id s
 \int \id r \,\Gamma(r-s) \,h_s\, F_r(x_r) +
 \cO(h^2)\label{de}
\end{multline}
\end{widetext}
to first order in the perturbation $h_t$. For stochastic integration
and path integrals in the non-Markovian case, see also \cite{han} and the more recent \cite{dyk} with additional references.
The equations \eqref{act}--\eqref{de} define our dynamical ensemble
for the path-space distribution with respect to the unperturbed
dynamics $\eqref{vin}$.  

Finally a word about initial and boundary conditions, also important
for the terminology.  As should be clear from the start, the
dynamics \eqref{vin} is not microscopic and its content depends on
chosen levels of description, including spatio-temporal scales.  In
the discussion of diffusion one assumes no spatial confinement over
the relevant time-scales.  The behavior is then transient concerning
the position-degrees of freedom, but beyond the inertial regime the
velocities relax and become Maxwellian.  In that regime overdamped
approximations can be valid which can be formally obtained in what
follows by putting $\dot{v}=0$.  For the general question of linear
response we also have in mind the case where the force $F_t = F + K_t$
contains a time-independent conservative force $F = -\nabla U$ from a
normalizable and confining potential $U$. 
Under such a confinement and for time-independent forcing $K_t=K$ it
is then assumed that there is a unique and smooth stationary density
$\rho(x,v)$ to which all initial data converge.  We speak of an
equilibrium dynamics when $K_t=0$ (only confining potential).  The
case of free diffusion $F_t=0$ is however not strictly equilibrium as
it needs not be stationary even in the velocity degrees of freedom. For
example for free diffusion the Sutherland-Einstein relation will only
be recovered when the initial data are also randomly chosen from a
Maxwellian.  In the formalism below, the important property of full
equilibrium will be stationarity combined with time-reversal
invariance.  In fact, to the dynamics \eqref{vin} must still be added
a relation between the noise correlations and the memory kernel, so as
to ensure for example that for $F_t=0$ the velocities become
Maxwellian.  The next section takes this to a more general discussion
by formulating the condition of local detailed balance.

\subsection{Entropy flux and local detailed balance}
The Einstein relation, also called the second fluctuation--dissipation
relation, connects the noise correlations to the memory kernel in the
friction.  For an equilibrium dynamics, say taking $F=h\equiv 0$ in \eqref{vin}, that relation (which will follow as \eqref{ld} below) is most simply derived from the requirement that the stationary velocity distribution should be Maxwellian at inverse temperature $\beta$.  For our nonequilibrium dynamics (non-conservative forcing $F$) we have in general no information about that stationary distribution.  Nevertheless, more fundamentally that Einstein relation arises in the weak coupling limit for the dynamical degrees of freedom with a large thermal bath as a consequence of the microscopic reversibility of the bath degrees of freedom.  For our dynamics we do keep the bath in thermal equilibrium and our driven particle does not react back on the bath. That is the basic reason why we will be able to maintain the standard Einstein relation.  As the situation with memory is formally more complicated and since we do not want to recall the details of a weak coupling analysis, we can rely on the so called condition of local detailed balance, see also \cite{har}.  It amounts to assuming that, if the system were subjected to
a confining force, it would reach equilibrium, and that the
exchanges with the thermostat are the same as when the exerted forces
drive it out of equilibrium. Formally, this means that the physical
entropy flux can be recovered from the time-antisymmetric part of the
action. That principle is indeed rooted in the reversibility of the
underlying microscopic dynamics as shown in \cite{mae03}.  To be more specific about time-reversal we introduce
the time-reversal operator $\theta$ according to which
\begin{eqnarray}
 \theta x_t & = & x_{-t},\quad
\theta v_t  =  -v_{-t}\nonumber\\
\theta h_t & = & h_{-t}, \quad \theta F_t = F_{-t}
\end{eqnarray}
where the last line also includes the time-reversal of the protocol
(the time-dependence in the non-magnetic external forces).  When the $\eta_t =(\eta_t^i)$ would be multidimensional we should also assume that the noise is time-reversal invariant in the sense that $\langle \eta^i_t\eta^j_0\rangle = \langle \eta^j_t\eta^i_0\rangle$.  Demanding that the
system obeys local detailed balance then means to require that
 the
time-antisymmetric part of the action for ${\cal P}_0$ (first appearing in \eqref{act}) is given by the
entropy flux in the original (unperturbed) model; that is to say
\begin{equation}\label{eq:entropy}
\log \frac{\id {\cal P}_0}{\id {\cal P}_0\theta}(\omega)= 
-\beta \int \id s \, \vdot_s\,v_s + \beta \int \id s \, F_s(x_s)\,v_s
\end{equation}
The first term in the right-hand side is a temporal boundary term
accounting for the kinetic energy difference between the initial and
final state of the trajectory. 

 As is explicitly shown in Appendix \ref{app:entropy},  local detailed balance \eqref{eq:entropy}
is verified whenever
\begin{equation}
\avg{\eta_s \eta_t} = \frac{1}{2} \left[ \gamma(t-s) + \gamma(s-t) \right]=
\frac{1}{2}\,\gamma(|t-s|)
\label{loc_det_bal}
\end{equation}
between the noise covariance and the symmetric part of the memory
kernel.  We repeat that \eqref{loc_det_bal} is as such independent of $F$ (nonequilibrium driving) or $h$ (perturbation) as it expresses the thermal equilibrium of the bath; it formally appears (as \eqref{eq:entropy}) from requiring
that the source of time-reversal breaking equals the
(excess) entropy flux.  
The very same condition \eqref{loc_det_bal} then also ensures that
\begin{equation}\label{ss}
 \cA_h(\theta\omega) - \cA_h(\omega) =
\beta \int \id s\, v_s \,h_s
\end{equation}
with right-hand side equal to the path-dependent excess entropy flux towards the
environment at inverse temperature $\beta$, or the dissipated power by
the force $h_t$ (setting $k_B=1$). 
 The identity \eqref{ss} is also
explicitly discussed in Appendix \ref{app:entropy}: we can derive \eqref{loc_det_bal} from requiring that the
time-antisymmetric part of the action ${\cal A}_h$ is the excess
entropy flux caused by the force $h_t$, thus equal to $\beta \int \id
s\, v_s \,h_s$.  We will use the
notation $\cS^{ex}(\omega) = \cA_h(\theta\omega) - \cA_h(\omega)$
for \eqref{ss} in what follows below.\\
It is worth to note  that
mathematically (\ref{loc_det_bal}) leads to simply
rewriting \eqref{del} as
\begin{equation}\label{ld}
\int \id s\, \Gamma(t-s)\, \gamma(|s|) = 2\,\delta(t)
\end{equation}

\subsection{The time--symmetric part, or the activity}
The time--symmetric part of the action ${\cal A}_h$ is
$\cT^{ex}(\omega) = \cA_h( \theta \omega) + \cA_h(\omega)$ and is
calculated to be
\begin{multline}
\label{traffic}
\cT^{ex}(\omega)  = \beta
\int \id r \,H_r \, (F_r(x_r)- \vdot_r) \\
+ \frac{\beta}{2}\;\int \id u \, v_u \int \id r\,H_r\left[ \gamma(u-r)
- \gamma(r-u) \right]
\end{multline}
where we have introduced the ``smeared-out'' perturbation
\begin{equation*}
H_r = \int \id s \,h_s \,\Gamma(r-s)
\end{equation*}
This time-symmetric part $\cT^{ex}$ is a function on path space and is also called the dynamical activity; it is related to the frenetic
contribution in linear response playing an important role when away from equilibrium, see \cite{up,fdr}.
Note that the antisymmetric part of the memory kernel vanishes in the Markov case. In this limiting case, $H_s =
h_s/\gamma$ and then
\[ 
\cT^{ex}_{\text{Markov}}(\omega) = \frac{\beta}{\gamma}\int \id s\,
h_s \,(F_s(x_s) - \vdot_s)
\] 
to linear order in $h_t$, as before.

\section{Linear response relations}\label{linex}

In what follows we denote by
$\langle \cdot \rangle_h, \langle \cdot \rangle $ the average over the
paths generated by \eqref{vin} with or without the perturbation $h_t$.
The only randomness over which we average is the stationary noise
$\eta_t$ but sometimes an additional average over initial conditions
will be mentioned.  For a general observable local in time we write
$O(x_t,v_t) = O_t$
\subsection{General susceptibility}
The linear response of observable $O$ is
obtained from
\begin{equation}
\label{eq:lin_resp}
\langle O_t\rangle_h - \langle O_t\rangle = -\langle O_t\,\cA_h\rangle
\end{equation}
or in terms of the generalized susceptibility $\chi_O$ defined
by:
\[ 
\chi_O(s,t) = \left. \frac{\delta}{\delta h_s}\langle O_t\rangle_h
\right|_{h=0}
\] 
From inserting \eqref{de} into \eqref{eq:lin_resp} we find
\begin{multline}\label{ch}
\chi_O(s,t) = \frac{\beta}{2}
 \int \id r \int \id u \,\Gamma(r-s)
\,\gamma(r-u)\,\avg{ O_t \,v_u } \\
+\frac{\beta}{2} \int \id r \,\Gamma(r-s) \,
\pnt{\avg{\vdot_r O_t}-\avg{F_r(x_r) O_t}}
\end{multline}
There are different ways to write that
same formula. We can add and subtract to get the symmetric part of the
memory kernel:
\begin{multline}
\chi_O(s,t) = \beta \langle O_t\,v_s\rangle
-\frac{\beta}{2} \int \id r \int \id u \,\Gamma(r-s)
\,\gamma(u-r)\,\avg{ O_t \,v_u } \\
 + \frac{\beta}{2}\int \id
 r \,\Gamma(r-s) \, \pnt{\avg{\vdot_r O_t}- \avg{F_r(x_r) O_t} }
\label{cordif}
\end{multline}
Another possibility, is to consider separately the time--antisymmetric
and the time--symmetric part. In that case we follow the decomposition of the action
$\cA_h=(\cT^{ex}-\cS^{ex})/2$, and from (\ref{eq:lin_resp}) the
susceptibility reads
\begin{equation}
\label{eq:fdtneq}
\chi_O(s,t)= \frac{1}{2} \avg{\sigma_s O_t} - \frac{1}{2}\avg{\tau_s O_t}
\end{equation}
where
\[ 
\sigma_s= \left. \frac{\delta}{\delta h_s} \cS^{ex} \right|_{h=0} =
\beta v_s
\] 
and
\begin{multline}
\label{eq:tau}
\tau_s=\left. \frac{\delta}{\delta h_s} \cT^{ex} \right|_{h=0} =
\beta \int \id r \, \Gamma(r-s)  (F_r- \vdot_r) \\+
\frac{\beta}{2}\;\int \!\! \id u  \int \!\! \id r \, v_u\, \Gamma(r-s) \left[ \gamma(u-r) - \gamma(r-u) \right]
\end{multline}
The formulation \eqref{eq:fdtneq} separates an entropic from a frenetic contribution as suggested e.g in \cite{fdr}.  The formul{\ae} \eqref{ch}--\eqref{cordif}--\eqref{eq:fdtneq} are the first principal resuls of the paper; they give a general understanding of the structure of nonequilibrium response also in the presence of memory effects. Moreover, there is the promise in each of the last terms in their right-hand sides to learn about the nonequilibrium driving exactly by the study of the response and especially from the frenetic contribution.\\

The deviation from the Markov case is felt only
in this excess dynamical activity $\cT^{ex}$ and not in
the excess entropy flux $\cS^{ex}$. That explains why only in
nonequilibrium situations the fluctuation-response relations change
when going from Markov to non-Markov; in equilibrium only the entropy
fluxes enter in fluctuation-response relations.  Of course, the
transient diffusive case is a nonequilibrium situation, and we should
be careful when possibly identifying $F_r=0$ with the equilibrium
case.

\subsection{Consequence of causality}
Causality requires that observations before a certain time are not
influenced by perturbations after that time.  As a consequence, from \eqref{eq:fdtneq},
\begin{equation}\label{ca}
 \avg{\sigma_u O_r} = \avg{\tau_u O_r}
 \end{equation}
for a time-ordering $u>r$.  But suppose now that the averages satisfy time-reversal invariance so that
\[
 \avg{\tau_s O_t} = \mbox{sgn }O\, \avg{\tau_{-s} O_{-t}},\quad \mbox{sgn }O\avg{\sigma_{-s} O_{-t}} = -\avg{\sigma_s O_t}
 \]
 where sgn $O$ is the parity of observable $O$ under time-reversal.
 Then, under that time-reversibility and as a result of the causality relation \eqref{ca} for $u=-s > r=-t$,
  \begin{eqnarray}
  \avg{\sigma_s O_t} - \avg{\tau_s O_t} &=& \avg{\sigma_s O_t} - \mbox{sgn } O\avg{\tau_{-s} O_{-t}}
\nonumber\\
&=& \avg{\sigma_s O_t} - \mbox{sgn } O\avg{\sigma_{-s} O_{-t}}
\\
&=& \avg{\sigma_s O_t} + \avg{\sigma_{s} O_{t}} = 2\beta\,\langle v_s O_t\rangle
\nonumber
\end{eqnarray}
which, upon inserting in \eqref{eq:fdtneq},  yields the standard fluctuation--dissipation relation
\begin{equation}
\label{eq:fdteq}
\chi_O(s,t) = \beta \avg{O_t \, v_s} \Theta(t-s)
\end{equation}
The next section comes back to this with yet another derivation.\\

Another consequence of causality is that the action \eqref{de} verifies
\[ 
\avg{O_t\, \cA_h} =0
\] 
when $h_s=0$ for $s\leq t$.  That immediately implies that for all $s > t$,
\begin{multline}\label{imm}
 \int  \id r \int \id u \,\Gamma(r-s)
\,\gamma(r-u)\,\avg{O_t \,v_u}\\  = \int \id r \,\Gamma(r-s) \,
\pnt{\avg{F_r(x_r) O_t} - \avg{\vdot_r O_t}}
\end{multline}
There is a simpler identity that applies when the original dynamics is
time-homogeneous, i.e., when $F_t=F$ does not explicitly depend on
time $t$.  Then we can think of the unperturbed averages $\langle
\cdot\rangle$ as a steady regime.  In that case we take time $t$ very
negative in \eqref{imm}, multiply both sides with
$\langle\eta_s\eta_w\rangle$ for arbitrary $w>t$, integrate over all
$s$ and use the identity \eqref{del} to obtain
\[ 
\int_t^{+\infty} \id u\,\gamma(w-u)\,\langle O_t\,v_u\rangle =
\avg{F(x_w) O_t} - \avg{\vdot_w O_t}
\] 
In the Markov case this identity is, for $w>t$:
\begin{equation}
\gamma \,\frac{\id}{\id w}\avg{ O_t\,x_w} =  \avg{F(x_w)\, O_t}
- \avg{\vdot_w \, O_t}
\end{equation}
which is readily recognized as $\avg{ O_t\,\eta_w} = 0$ for white noise $\eta_w, w>t$.

\subsection{Equilibrium dynamics}
\subsubsection{Confined case}
The equilibrium limiting case can be achieved whenever the force field
$F$ derives from a potential function. In that case time-reversal
invariance applies, and one should recover from (\ref{eq:lin_resp})
the standard fluctuation--dissipation theorem \eqref{eq:fdteq}. To check this more explicitly, we first
consider the response of an observable $O_t \equiv O(x_t,v_t)$ which
is {\it even} under time--reversal (i.e. $\theta O_t = O_{-t}$).  In
this case
\begin{eqnarray}
\label{eq:timereversal}
\avg{ O_t\,v_{-u}} & =-\avg{ O_{-t}\,v_{u}} \nonumber\\
\avg{F_r(x_r)\,O_{-t}} &=\avg{F_{-r}(x_{-r})\,O_{t}}\\
\avg{\vdot_r\,O_{-t}} &=\avg{\vdot_{-r}\,O_{t}} \nonumber
\end{eqnarray}
In
Appendix \ref{app:equilibrium} we show that the time anti--symmetric part of the response
$\chi$ yields
\begin{equation}
\label{eq:fdteq2}
\chi_O(s,t)-\chi_O(t,s)= \beta \avg{O_t \, v_s}
\end{equation}
By causality, this automatically implies the
fluctuation-dissipation relation \eqref{eq:fdteq}.\\

Likewise, when the observable $O_t$ is odd under time reversal
(i.e. $\theta O_t=-O_{-t}$), the right-hand sides in \eqs
(\ref{eq:timereversal}) have to be multiplied by $-1$, and thence one
has to consider the time--symmetric part $\chi_O(s,t)+ \chi_O(t,s)$ in
order to indeed recover (\ref{eq:fdteq}).

\subsubsection{Free diffusion}
We next consider the case where $F_r=0$ and there is no confinement on the relevant time-scales. In this case the
velocity of the particle relaxes alright to a Maxwellian steady--state in the long time limit, but its position diffuses. (It could be anomalous diffusion for
slowly decaying kernels.)  Therefore, starting from a fixed position and Maxwellian velocity, the velocity response $\chi_v$ satisfies
\begin{equation}
\label{eq:mobility}
\chi_v(s,t) = \chi_v(0,t-s) = \beta \avg{v_s v_t}
\end{equation}
for $s<t$, but the position dynamics remains in the transient regime. It is however possible to recover a formula similar to (\ref{eq:fdteq}) which
involves the mean square displacement $\Delta
x^2(t)=\avg{(x_{s+t} - x_s)^2}$ as we now explain.\\
 We consider the case
where the observable $O$ is the position $x$.  The position response 
\[\chi_x(t) = \left. \frac{\delta}{\delta h}\langle x_t\rangle_h
\right|_{h=0} = \int_0^t \id s \,\left. \frac{\delta}{\delta h_s}\langle x_t\rangle_h
\right|_{h=0}  
\]
satisfies
\begin{equation}
\label{eq:compmob}
 \frac{\id}{\id t} \chi_x(t) = \int_0^t\id s\,\chi_v(s,t) = \beta\,\int_0^t\id s\,\avg{v_s v_t}
\end{equation}
On the other hand,
\begin{eqnarray*}
\Delta x^2(t) &=& \int_s^{s+t} \id u  \int_s^{s+t} \id r
\avg{v_{u} v_{r}}\\
\frac{\id}{\id t} \Delta x^2(t) &=& 2 \int_s^{s+t} \id u \avg{v_{u} v_{s+t}}
\end{eqnarray*}
As a result, we get
the equilibrium-like result
\begin{equation}\label{chit}
\chi_x(t) = \frac{\beta}{2} \frac{\id}{\id t} \Delta x^2(t),\quad t>0
\end{equation}
which, for the Markov case, is an identity attributed to Virasoro in \cite{ckp}.  The fact that this relation remains satisfied also in the presence of strong memory is directly caused by $F=0$ and that only the entropic contribution matters.

\subsection{Modified Sutherland-Einstein relation}\label{sein}
We now show how to connect the mobility of the particle (which is related
to the velocity response to a constant force), and the diffusion
properties, related to the time behaviour of the mean squared
displacement. Remember that the perturbing field is a step
function $h_t=h \Theta(t)$, with constant field $h$. The
time--dependent mobility is then defined as:
\begin{equation}
M(t)=\frac{1}{t} \left. \frac{\partial}{\partial
h} \avg{(x_t-x_0)}_h \right|_{h=0} 
\end{equation}
or
\begin{equation}\label{mt}
M(t)=\frac{1}{t} \int_0^t \id s \int_0^t \id r \, \chi_v(s,r)
\end{equation}
The time--dependent diffusion coefficient is defined as
\begin{equation}
\label{eq:diffusion}
\tilde D(t)=\frac{1}{2 t} \,\Delta x^2(t) =\frac{1}{2 t} \int_0^t \id
r \int_0^t \id s \avg{v_s v_r}
\end{equation}
From the general linear response  (\ref{eq:fdtneq}) we know that
\begin{equation}
\label{eq:vcorr}
\avg{v_s v_t} = \frac{2}{\beta} \chi_v(s,t) + \frac 1{\beta} \avg{\tau_s v_t}
\end{equation}
which can be replaced in \eqref{mt}--(\ref{eq:diffusion}) to give a relation between $M$,
$\tilde D$ and $\tau$,
\[ 
M(t)=\beta \tD(t) - \frac{1}{2t} \int_0^t \id r \int_0^t \id
s \avg{\tau_s v_r}
\] 
We see how the violation of the Sutherland--Einstein relation $M=\beta \tilde D$ is related to the time-averaged correlation between displacement and dynamical activity,
\begin{eqnarray}\label{mse}
M(t) &=& \beta \tD(t) - \frac{1}{2t} \int_0^t \id
s \avg{\tau_s (x_t-x_0)} \\
&=&
\beta \tD(t) - \frac{1}{2} \avg{(x_t-x_0)\,\frac 1{t}\int_0^t \id s\, \tau_s }\nonumber
\end{eqnarray}
This formula has a general validity, independent of initial
conditions.  The time-averaged dynamical activity due to the
perturbation has become here an important observable for describing
the deviation from the standard Sutherland--Einstein relation. Since
$\tau_s$ is time-symmetric, the last correction term will vanish when
time-reversal symmetry gets established; that happens in the case of

free diffusion upon averaging over the initial equilibrium
distribution for velocities (as in the previous section).  In general
however we can further detail the expression by using the
explicit form of $\tau_s$ from (\ref{eq:tau}), yielding three terms
in the correction:
\begin{equation}\label{msee}
M(t)= \beta \tD(t) + \tC_1(t) + \tC_2 (t) + \tC_3(t) \,\,.
\end{equation}
As we are however most interested in the nonequilibrium situation with
possible drift, it is useful to discard this effect by
considering a slighly different definition of the diffusion
coefficient, relabeled here as $D$:
\begin{equation}
 D(t) = \frac{1}{2t} \langle (x_t - x_0); (x_t - x_0)\rangle.
\end{equation}
The notation $\langle A;B \rangle$ refers to the connected (also called truncated) correlation
function $\langle AB \rangle-\langle A \rangle\langle B \rangle$.
The analogue of the modified Sutherland--Einstein relation (\ref{msee}) 
is explicitly given by:
\begin{widetext}
	 \begin{eqnarray}
&&M(t) =  \beta D(t) +\underbrace{ \frac{\beta}{2t}\int_0 ^{t} ds\,\int \id r \,\Gamma(r-s)\,\langle (x_t-x_0) ;\,\dot{v_r}\rangle}_{ C_1(t)}
\nonumber\\
\!\!\!\!\!\!\!\!\!\!\!\!\!\!\!\!\!\!\!\!\!\!	+ &&\underbrace{ \frac{\beta}{4t}\int_0 ^{t} ds[ \int \id r \int \id u \Gamma(r-s)\{\gamma(r-u)-\gamma(u-r)\}\,\langle (x_t-x_0) ;v_u \rangle]}_{ C_2(t)} 
 \underbrace{-\frac{\beta}{2t}\int_0 ^{t} \int \id r \,\Gamma(r-s) \, \langle F_r(x_r); (x_t-x_0)\rangle\;ds}_{ C_3(t)}\nonumber \,\,.\\
\label{es}
			  \end{eqnarray}
\end{widetext}
This is the third key relation of the present paper (after the general \eqref{ch}--\eqref{cordif}--\eqref{eq:fdtneq} and the equilibrium-like \eqref{chit}).
We have postponed a detailed derivation of this result in
Appendix \ref{app:truncated}. Notice that the functions $\tC_i$ of \eq
(\ref{msee}) are simply given by the functions $C_i$ with a standard
correlation function instead of a connected one. More concretely, if
$C_i = \avg{A;B}$ then $\tC_i= \avg {A B}$. We shall further explore
this relation numerically in the next section. The terms $C_1$ and
$C_3$ are also present (although without convolution with $\Gamma$) in
the Markov case, and accounts for nonequilibrium effects due
respectively to the inertia and to the nonequilibrium forcing. The
term $C_1$ vanishes outside the inertial regime.  The term $C_3$ is
most important for the inverse problem of reconstructing the
nonequilibrium driving from violation of the Sutherland--Einstein
relation. The term $C_2$ has no analogue in the Markovian case, and is
hence only due to memory effects under nonequilibrium dynamics.

\subsection{Examples}\label{secse}
\begin{figure}
\begin{center}
  \def\svgwidth{0.9 \columnwidth} 
  \executeiffilenewer{one_cf_f0scale.svg}{one_cf_f0scale.pdf}%
  {inkscape -z -D --file=one_cf_f0scale.svg %
  --export-pdf=one_cf_f0scale.pdf --export-latex}%
  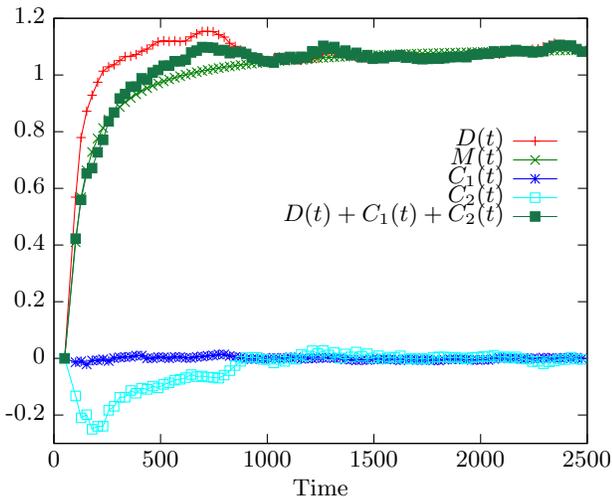%

\end{center}
\caption {The time-dependence of the mobility $M$, the diffusion $D$ and the corrections $C_1$ and $C_2$ of \eqref{es} after multiplying all with $\log t$.  Here we have long memory $\gamma(t) = \frac{1}{1+t}$ and free diffusion $F=0$ at  $\beta=1$.}
\label{fig1}
\end{figure}

\begin{figure}
\begin{center}
  \def\svgwidth{0.9 \columnwidth}
  \executeiffilenewer{sq_cf_f0scale.svg}{sq_cf_f0scale.pdf}%
  {inkscape -z -D --file=sq_cf_f0scale.svg %
  --export-pdf=sq_cf_f0scale.pdf --export-latex}%
  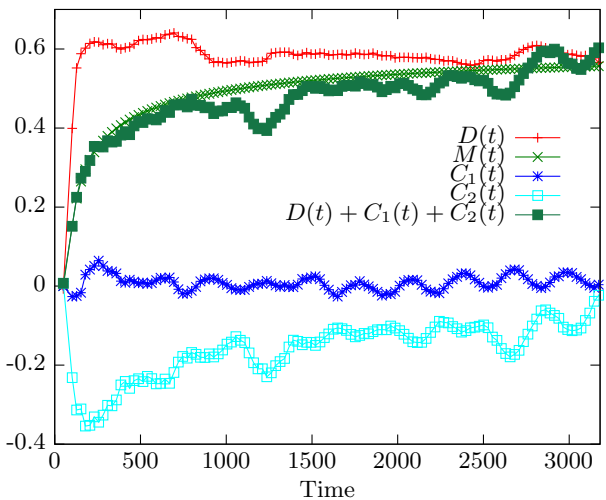%

\end{center}
\caption {Same case of free diffusion as in Fig. \ref{fig1} but with a time-rescaling of $\sqrt{t}$ for memory kernel $\gamma(t) = \frac{1}{\sqrt{1+t}}$. Still $F=0, \beta=1$.}
\label{fig2}
\end{figure}

\begin{figure}
\begin{center}
 \def\svgwidth{0.9 \columnwidth} %
  \executeiffilenewer{rfD.svg}{rfD.pdf}%
  {inkscape -z -D --file=rfD.svg %
  --export-pdf=rfD.pdf --export-latex}%
  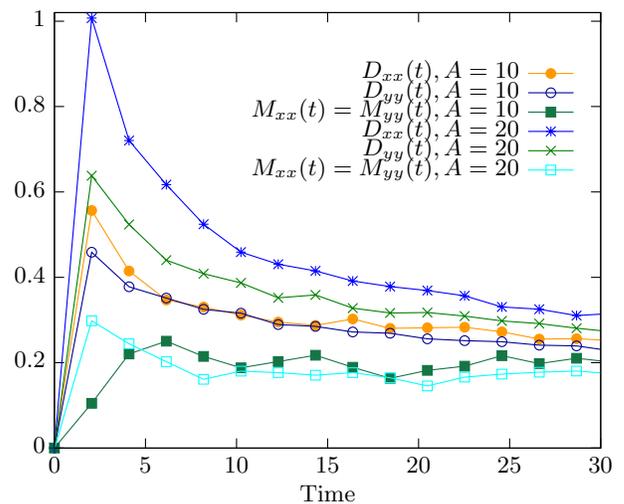%

\end{center}
\caption {Diffusion and mobility in the case of a two-dimensional rotational forcing \eqref{2dim} for memory $\gamma(t) = \frac{1}{1+t}$ and amplitudes $A=10 , A=20$.}
\label{fig3}
\end{figure}

\begin{figure}
\begin{center}
  \def\svgwidth{0.9 \columnwidth}
  \executeiffilenewer{one_cw_f1x5_1_1scale.svg}{one_cw_f1x5_1_1scale.pdf}%
  {inkscape -z -D --file=one_cw_f1x5_1_1scale.svg %
  --export-pdf=one_cw_f1x5_1_1scale.pdf --export-latex}%
  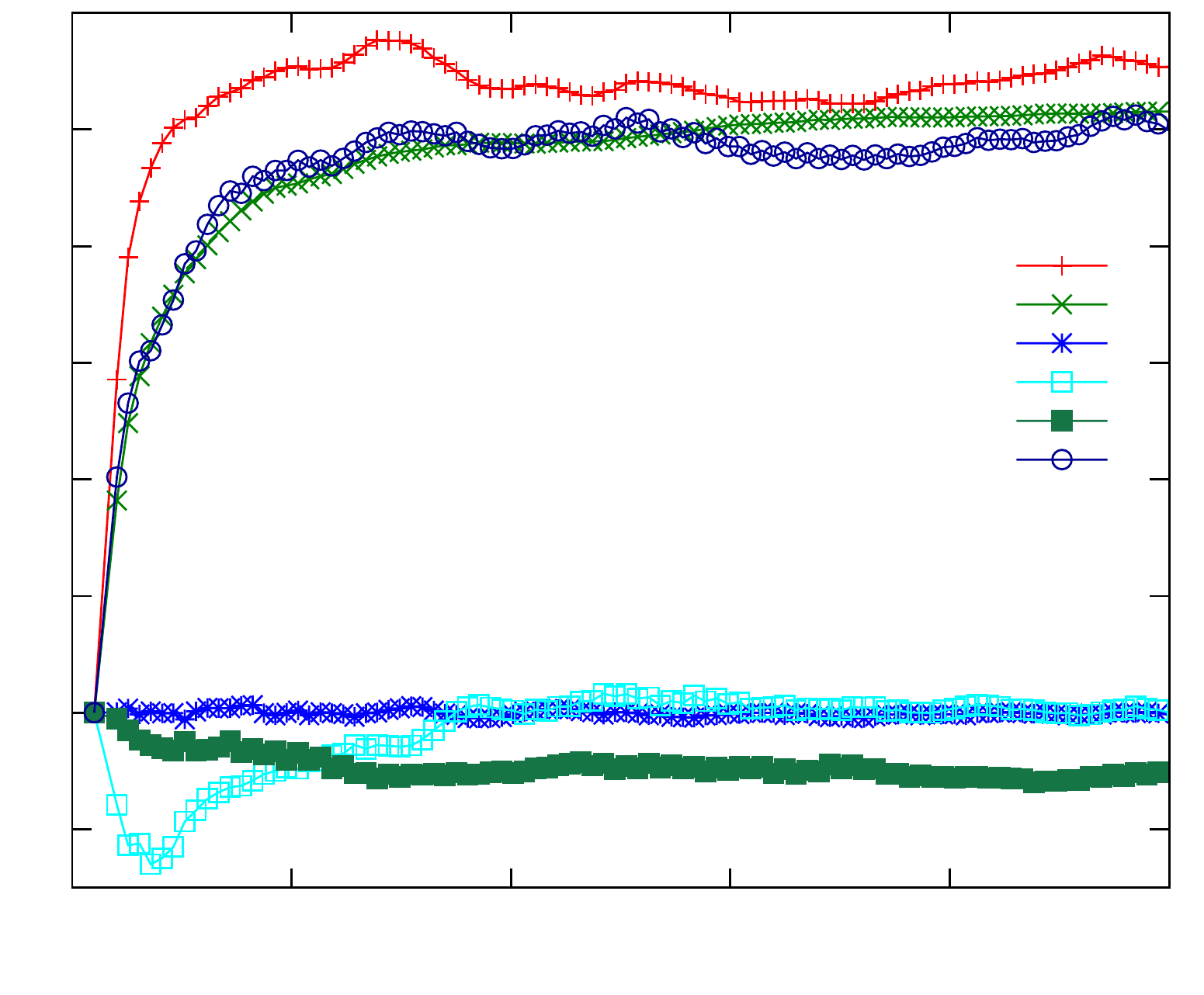%

\end{center}
\caption {Same set-up as in Fig.~\ref{fig1} but for a forcing $F(x)=1.5+\sin x$ (external field over periodic potential) at inverse temperature $\beta=1$. The rescaling of the mobility, the diffusion and the corrections in \eqref{es} is by multiplying all with $\log t$ for memory kernel $\gamma(t) = \frac{1}{1+t}$.}
\label{fig4}
\end{figure}

\begin{figure}
\begin{center}
  \def\svgwidth{0.9 \columnwidth}
  \executeiffilenewer{sq_cw_f1x5_1_1scale.svg}{sq_cw_f1x5_1_1scale.pdf}%
  {inkscape -z -D --file=sq_cw_f1x5_1_1scale.svg %
  --export-pdf=sq_cw_f1x5_1_1scale.pdf --export-latex}%
  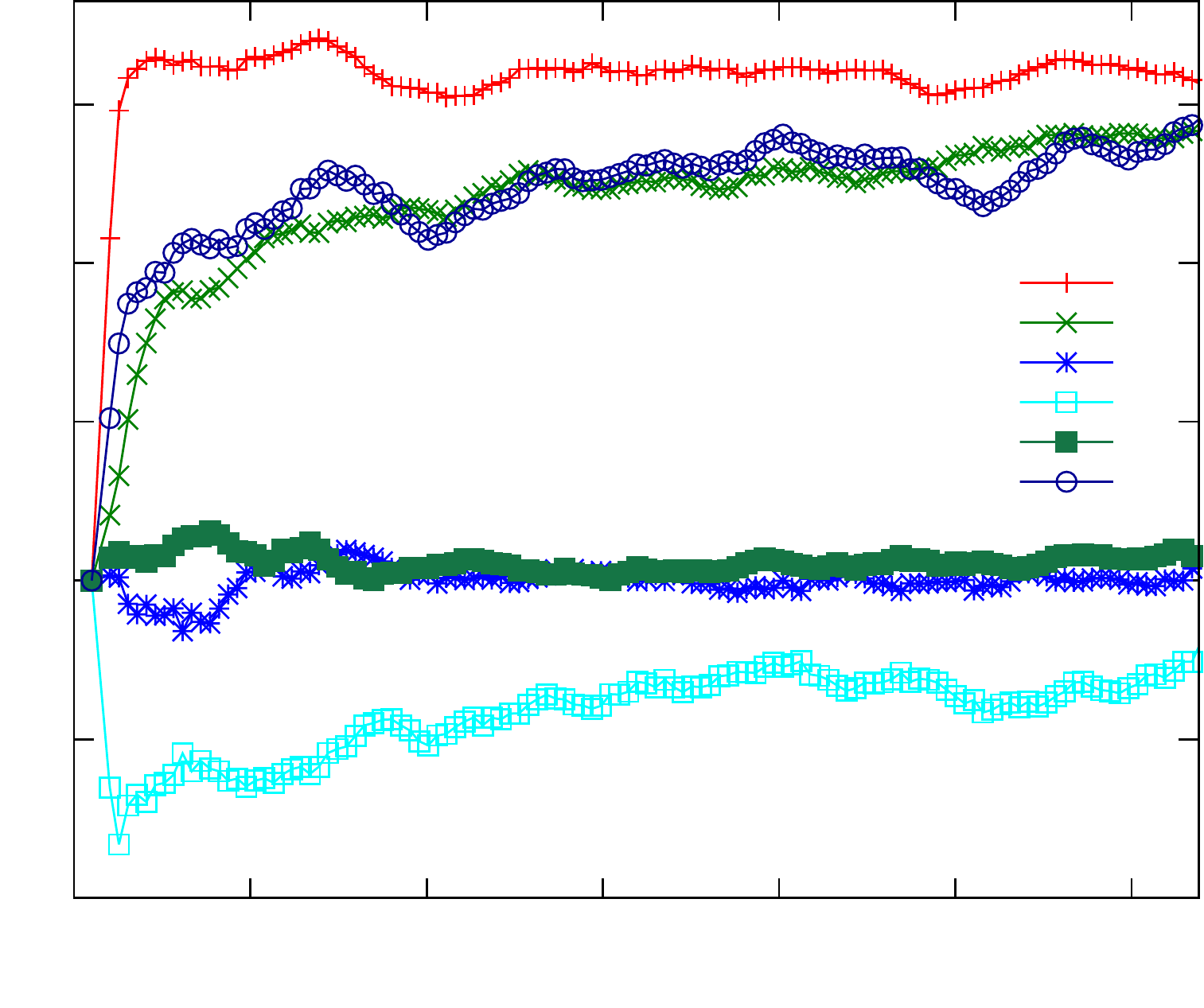%

\end{center}
\caption {Same set-up as in Fig.~\ref{fig2} but for a forcing $F(x)=1.5+\sin x$ (external field over periodic potential) at inverse temperature $\beta=1$. The rescaling of the mobility, the diffusion and the corrections in \eqref{es} is by multiplying all with $\sqrt{t}$ for memory kernel $\gamma(t) = \frac{1}{\sqrt{1+t}}$.}
\label{fig5}
\end{figure}

\begin{figure}
\begin{center}
  \def\svgwidth{0.9 \columnwidth}
  \executeiffilenewer{sq_cw_f8_8_1scale.svg}{sq_cw_f8_8_1scale.pdf}%
  {inkscape -z -D --file=sq_cw_f8_8_1scale.svg %
  --export-pdf=sq_cw_f8_8_1scale.pdf --export-latex}%
  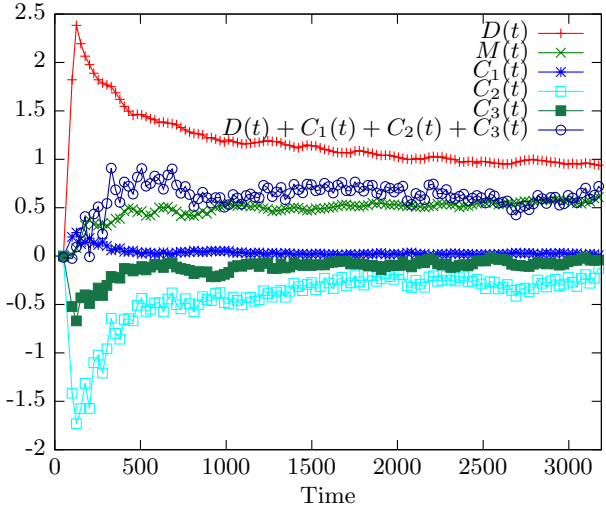%

\end{center}
\caption { The rescaling of the mobility, diffusion and the corrections by multiplying with $\sqrt{t}$ for $\gamma(t) = \frac{1}{\sqrt{1+t}}$, and $F(x)=8+8\sin x, \beta=1$.}
\label{fig6}
\end{figure}

We present here simulation results for the dynamics \eqref{vin} for some three choices of the driving $F$.   The method to generate colored noise is outlined in Appendix \ref{sim}.  For the evaluation of the terms in \eqref{es} we note that $\Gamma$ appears as a convolution and the Fourier transform of $\Gamma$ is computed from its definition \eqref{del}.  \\ 
  The purpose is to visualize the various terms in the modified Sutherland-Einstein relation \eqref{es}, similarly to the work in \cite{proc} but with the extra ingredient of memory. Of course there is no strong need to verify or to confirm the mathematical formula \eqref{es} as the linear expansion is sufficiently controlled, but it is somewhat interesting to see the contributions of the various terms in \eqref{es} and to see what influences them.\\
    We start with the free diffusion ($F=0$).
 The system is diffusive if, for large times, $D(t)$ reaches a limit (the diffusion constant) $D$.
 On the other hand, when the memory gets long time tails, with $\gamma(t)$ decaying algebraically, the diffusion can become anomalous, \cite{noe}.
We take two different examples, $\gamma(t)=1/(1+t)$ and
$\gamma(t)=1/\sqrt{1+t}$.  As results show, both the diffusion and the mobility follow a corresponding temporal behavior, with for the first example
 $D(t)\simeq 1/\log t$ and for the second example
$D(t) \simeq 1/\sqrt{t}$  (subdiffusive motion). The plots in Fig.\ref{fig1} and  Fig.\ref{fig2} show the subdiffusive behaviour due to the memory effect. More generally for free diffusion $D\propto (\int_0^{t}\gamma(s) \,\id s)^{-1} $ for large $t$.   Note that we treated free diffusion  with some fixed initial condition (without initial velocity-averaging) so that the standard Sutherland-Einstein relation gets established only in the long time limit.  Note that the inertial regime is rather short-lived, $C_1$ rapidly being very small, but the memory effect as present in $C_2$ postpones the equality  between $\beta D(t)$ and $M(t)$ to longer times.  The sign of $C_2$ carries no special information.\\
For the other examples we switched on some rotational force $F$.  Again we checked in all cases that for exponentially decaying memory kernels $\gamma(t)$ the results of \cite{proc} are reproduced.  We concentrate on power law decay and we consider finite times.
In Fig.\ref{fig3}  we consider a vector force on the plane to induce vortices, similar to the rotational force in \cite{proc}. We take $\vec{F} = A \vec{g}$, with $A$ the amplitude,  where
\begin{eqnarray}
 g_x(x,y)&=& a(r-\sqrt2)(y-\frac{1}{2})\label{2dim}\\
g_y(x,y)&=& a(r-\sqrt2)(\frac{1}{2}-x)\nonumber\\
a&=& (1-2\delta_{2,x\,\mathrm{mod}\,3})(1-2\delta_{1,y\,\mathrm{mod}\,2})\nonumber
\end{eqnarray}
for distance $r=\sqrt{(x-\frac{1}{2})^2+(y-\frac{1}{2})^2}$.  The somewhat involved definitions assure that the particle does not undergo a net drift; the forces are purely rotational and not translational now.
To make sure, mobility and diffusion are now matrices but the off-diagonal elements are approximately zero.  We also find that the diffusion in the $x$-direction is bigger than in the $y$-direction. Moreover, for bigger $A$, the diffusion increases, while the mobility remains almost constant
(and even somewhat decreases).\\
Finally, we consider the result of driving in 1 dimension. The nonequilibrium force is obtained from a periodic potential, which is like confining the particle to a toroidal trap, and adding a constant field. In formul{\ae}, the
nonequilibrium force is $F(x) =A+\sin x$, where $A$ is a
constant. 
Fig.\ref{fig4} and  Fig.\ref{fig5} show the result again for the two long-memory kernels.
As we have found in the previous section the relation between diffusion and mobility gets modified.
Our simulations confirm in all  cases that the diffusion depends on the external forcing more strongly than does the mobility.  
In the power law decaying memory for $A=1.5$ the increase of the diffusion is seen.   $C_1$ is still almost zero, $C_2$ vanishes in the long time and the diffusion and mobility are not proportional any more;
there is now also the essential term $C_3$, which is negative because of the positive correlation between force and displacement. However, for stronger memory and with $A=1.5$ there is little difference with the case $f=0$. To make the contribution of $C_3$ more prominent we have taken a larger driving force, like $F(x) =8+8\sin x $ as in Fig.\ref{fig6}; we see
that $C_1$ is still zero, $C_2$ is getting smaller faster than before and $C_3$ is more important now. \\




\pagebreak

\section{Relevance and conclusion}
The fluctuation-dissipation theorem has an extension to
nonequilibrium generalized Langevin systems which preserves the splitting of the response in an entropic and a frenetic contribution. That general statement is of course relevant in today's search for useful linear response formula away from equilibrium. The presence of memory is especially relevant for dense colloidal suspensions.  We have here considered a driving which can be time-inhomogeneous or even random but we have also assumed the presence of Gaussian (correlated) noise to start an expansion from path-integrals for an underdamped dynamics.  The Gaussian correlations are in fact connected with the memory kernel in the friction via the condition of local detailed balance.   An important application is to the extension of the Sutherland-Einstein relation between diffusion and mobility.  Because of the nonequilibrium condition, the diffusion constant is no longer alone in determining the transport properties of colloidal particles.   That was in particular seen in  detailed simulations
for various nonequilibrium diffusions, possibly anomalous, in particular for exploring the
role of the nonequilibrium forcing and the influence of memory.  An interesting conclusion is that the nonequilibrium corrections to the Sutherland-Einstein relation are related to the time-correlations between the so called dynamical activity and the velocity of the particle, which in turn leads to information about the correlations between the driving force and the particle's displacement, \cite{pro}.  Such an analysis provides a more general framework for discussions on the violation of the fluctuation-dissipation theorem and is an alternative for the use of effective temperatures.  We believe that both the direct question (predicting the response) as the inverse question (deriving information on the nonequilibirum forcing) can be attacked within the given formalism.\\


\noindent {\bf Acknowledgment}  C.M. is grateful to Matthias Kr\"uger for interesting discussions and acknowledges the hospitality in the Condensed Matter Theory group at MIT.

\appendix

\section{Calculation of the entropy fluxes}
\label{app:entropy}

We start with deriving \eqref{eq:entropy},
From writing out the action $\log \frac{\id {\cal P}_0}{\id {\cal P}_0\theta}(\omega)$ in the same way as for \eqref{de} we find that
\begin{widetext}
\begin{equation}
\label{AA1}
 \begin{split}
\log \frac{\id {\cal P}_0}{\id {\cal P}_0\theta}(\omega) =  &- \frac{\beta}{2} \int ds
  \int dr \,\Gamma(r-s)\,\vdot_r
  \int du \,[\gamma(r-u)+ \gamma(u-r)]\,v_u\,
  \\ &- \frac{\beta}{4} \int ds
  \int dr \,\Gamma(r-s) \,\int
  du \,\gamma(s-u)\,v_u \int
  dw\,\gamma(r-w)\,v_w \\ &+ \frac{\beta}{4}
  \int ds \int dr
  \,\Gamma(r-s) \,\int du
  \,\gamma(u-s)\,v_u\int
  dw\,\gamma(w-r)\,v_w \\ &+  \frac{\beta}{2}
  \int ds \int dr
  \,\Gamma(r-s) \, F_r(x_r)\, \int du
  \,[\gamma(r-u)+ \gamma(u-r)]\,v_u
 \end{split}
\end{equation}
which can be rewritten as
\begin{equation}
\label{AA2}
\begin{split}
 = & -\frac{\beta}{2} \int ds
  \int dr \,\Gamma(r-s)\,\vdot_r
  \int du \,[\gamma(r-u)+ \gamma(u-r)]\,v_u\,
  \\ &  -  \frac{\beta}{4} \int ds
  \int dr \,\Gamma(r-s) \,\int
  du \,\gamma(s-u)\,v_u \int
  dw\,\gamma(r-w)\,v_w \\&  -  \frac{\beta}{4}
  \int ds \int dr
  \,\Gamma(r-s) \,\int du \,\gamma(u-s)\,v_u
  \int dw\,\gamma(r-w)\,v_w \\  &+
  \frac{\beta}{4} \int ds \int
  dr \,\Gamma(r-s) \,\int du \,\gamma(u-s)\,v_u
  \int dw\,\gamma(r-w)\,v_w \\   &+
  \frac{\beta}{4} \int ds \int
  dr \,\Gamma(r-s) \,\int du
  \,\gamma(u-s)\,v_u\int
  dw\,\gamma(w-r)\,v_w \\ & +  \frac{\beta}{2}
  \int ds \int dr
  \,\Gamma(r-s) \, F_r(x_r)\, \int du
  \,[\gamma(r-u)+ \gamma(u-r)]\,v_u
 \end{split}
\end{equation}
from which we arrive at
\begin{eqnarray}
 \label{AA3}
\log \frac{\id {\cal P}_0}{\id {\cal P}_0\theta}(\omega)&=& -\frac{\beta}{2} \int ds
  \int dr \,\Gamma(r-s)\,\vdot_r
  \int du \,[\gamma(r-u)+ \gamma(u-r)]\,v_u\,
  \nonumber\\ &-& \frac{\beta}{4} \int ds
  \int dr \,\Gamma(r-s) \,\int
  du \, \int
  dw\,[\gamma(s-u)+\gamma(u-s)]\gamma(r-w)\,v_w\,v_u\nonumber\\ &+&
  \frac{\beta}{4} \int ds \int
  dr \,\Gamma(r-s) \,\int du \,
  \int
  dw\,[\gamma(r-w)+\gamma(w-r)]\gamma(u-s)\,v_w\,v_u\nonumber\\ &+&
  \frac{\beta}{2} \int ds \int
  dr \,\Gamma(r-s) \, F_r(x_r)\, \int du
  \,[\gamma(r-u)+ \gamma(u-r)]\,v_u
 \end{eqnarray}
\end{widetext}
Since $\left< \eta_s \eta_t \right> = \frac{1}{2} \left[ \gamma(t-s) +
  \gamma(s-t) \right]$, as implied by local detailed balance \eqref{loc_det_bal}, and by
using the definition of the symmetric kernel $\Gamma(t)$, the second
term and the third term of the right-hand side cancel each other and one finally
recovers \eqref{eq:entropy}.\\
  Likewise, when we add a time-dependent perturbation $h_t$ we have
\begin{widetext}
\begin{eqnarray}
\label{S^h}
\log \frac{\id {\cal P}_h}{\id {\cal P}_h\theta}(\omega) &=& -\frac{\beta}{2} \int ds
  \int dr \,\Gamma(r-s)\,\vdot_r
  \int du \,[\gamma(r-u)+ \gamma(u-r)]\,v_u\,
  \nonumber\\ &-& \frac{\beta}{4} \int ds
  \int dr \,\Gamma(r-s) \,\int
  du \,\int
  dw\,[\gamma(s-u)\gamma(r-w)-\gamma(u-s)\gamma(w-r)]v_u\,v_w\nonumber\\ &+&
  \frac{\beta}{2} \int ds \int
  dr \,\Gamma(r-s) \, F_r(x_r)\, \int du
  \,[\gamma(r-u)+ \gamma(u-r)]\,v_u\, \nonumber\\ &+& \frac{\beta}{2}
  \int ds \int dr
  \,\Gamma(r-s) \, h_s\, \int dw [\gamma(w-r)+
    \gamma(r-w)]\,v_w\, \nonumber\\
 \end{eqnarray}
The excess entropy flux is then
\begin{eqnarray*}
 \cS^{ex}(\omega) &=&  {\cal A}_h(\theta\omega) - {\cal A}_h(\omega) =   \log \frac{\id {\cal P}_h}{\id {\cal P}_h\theta}(\omega) -  \log \frac{\id {\cal P}_0}{\id {\cal P}_0\theta}(\omega)\\
 &=& \frac{\beta}{2}
 \int ds \int dr \,\Gamma(r-s)
 \, h_s\, \int dw [\gamma(w-r)+
   \gamma(r-w)]\,v_w\,
\end{eqnarray*}
\end{widetext}
which indeed ensures \eqref{ss} upon using \eqref{ld}.

\section{Recovery of the equilibrium fluctuation--dissipation theorem}
\label{app:equilibrium}

We show how to get from (\ref{eq:lin_resp}) to (\ref{eq:fdteq2})
when (\ref{eq:timereversal}) holds. We start by writing

\begin{multline}
\chi_O(s,t)-\chi_O(t,s)  =\frac{\beta}{2}
 \int \id r \int \id u\, \Gamma(r-s) \gamma(r-u) \avg{O_t v_u}\\
+ \frac{\beta}{2} \int \id r \Gamma(r-s) \pnt{\avg{\vdot_r
O_t}- \avg{F_r O_t}} \\  -\frac{\beta}{2} \int \id r \int \id
u \Gamma(r-t) \gamma(r-u) \avg{O_s v_{u}} \\  -\frac{\beta}{2}\int \id
r \Gamma(r-t)
\pnt{\avg{\vdot_r
O_s}- \avg{F_r O_s}}
\end{multline}
In the last two integrals we can perform the change of variable
$r'=t+s-r$ and $u'=t+s-u$. By relabeling $r'=r$ and $u'=u$ one gets
\begin{multline}
\chi_O(s,t)-\chi_O(t,s) = \\
\frac{\beta}{2} \int \id r \int \id u \Gamma(r-s) \gamma(r-u) \avg{O_t
  v_u} \\- \frac{\beta}{2} \int \id r \int \id u \Gamma(r-s)
\gamma(u-r) \avg{O_s v_{t+s-u}} \\ + \frac{\beta}{2} \int \id r
\Gamma(r-s) \left[ \avg{\vdot_r
O_t}- \avg{F_r O_t} \right. \\ - \left. \avg{\vdot_{t+s-r}
O_s}+ \avg{F_{t+s-r} O_s}   \right]
\end{multline}
Finally, by using the time--translation invariance of correlation
functions plus the time--reversal conditions (\ref{eq:timereversal})
one will observe that the square brackets term in the last integral vanish
and the first two integrals simplify thanks to (\ref{loc_det_bal}) and
yields to \eq (\ref{eq:fdteq2}).

\section{The modified Sutherland-Einstein relation in connected form}
\label{app:truncated}
By inserting the dynamical activity in \eqref{mse} we obtain
\begin{widetext}
     \begin{eqnarray} M(t) &=&\frac{\beta}{2t}\langle
(x_t-x_0)^2 \rangle +\frac{\beta}{2t}\int_0 ^{t} ds\,\int \id
r \,\Gamma(r-s)\,\langle (x_t-x_0) \,\dot{v_r}\rangle\nonumber\\
&+& \frac{\beta}{4t}\int_0 ^{t} ds \int \id r \int \id
u \Gamma(r-s)\{\gamma(r-u)-\gamma(u-r)\}\,\langle (x_t-x_0)\,
v_u \rangle \nonumber\\ &-&\frac{\beta}{2t}\int_0 ^{t} \int \id
r \,\Gamma(r-s) \, \langle F_r(x_r)\,
(x_t-x_0)\rangle\;ds\nonumber \end{eqnarray}
\end{widetext}
We now rewrite this relation as follows
\begin{widetext}
     \begin{eqnarray}
M(t) &=&  \frac{\beta}{2t}\langle (x_t-x_0)^2 \rangle - \frac{\beta}{2t} \langle x_t-x_0 \rangle\,\langle x_t-x_0 \rangle +  \frac{\beta}{2t} \langle x_t-x_0 \rangle\, \langle x_t-x_0 \rangle \nonumber\\
&+&\frac{\beta}{2t}\int_0^{t} ds\,\int \id r \,\Gamma(r-s)\,\langle (x_t-x_0) \,\dot{v_r}\rangle\,
-\frac{\beta}{2t}\int_0 ^{t} ds\,\int \id r \,\Gamma(r-s)\,\langle x_t-x_0 \rangle\,\langle \dot{v_r}\rangle\nonumber\\
&+&\frac{\beta}{2t}\int_0 ^{t} ds\,\int \id r \,\Gamma(r-s)\,\langle x_t-x_0 \rangle\,\langle\dot{v_r}\rangle\nonumber\\
&+& \frac{\beta}{4t}\int_0 ^{t} ds \int \id r \int \id u \Gamma(r-s)\{\gamma(r-u)-\gamma(u-r)\}\,\langle (x_t-x_0) \, v_u \rangle \nonumber\\
&-& \frac{\beta}{4t}\int_0 ^{t} ds \int \id r \int \id u \Gamma(r-s)\{\gamma(r-u)-\gamma(u-r)\}\,\langle x_t-x_0 \rangle\,\langle v_u \rangle \nonumber\\
&+& \frac{\beta}{4t}\int_0 ^{t} ds \int \id r \int \id u \Gamma(r-s)\{\gamma(r-u)-\gamma(u-r)\}\,\langle x_t-x_0 \rangle\,\langle v_u \rangle \nonumber\\
&-&\frac{\beta}{2t}\int_0 ^{t} \int \id r \,\Gamma(r-s) \, \langle F_r(x_r)\, (x_t-x_0)\rangle\;ds\,
+\frac{\beta}{2t}\int_0 ^{t} \int \id r \,\Gamma(r-s) \, \langle F_r(x_r)\rangle\,\langle x_t-x_0 \rangle\;ds\nonumber\\
&-&\frac{\beta}{2t}\int_0 ^{t} \int \id r \,\Gamma(r-s) \, \langle F_r(x_r)\rangle\,\langle x_t-x_0 \rangle\;ds\nonumber\\
	 \end{eqnarray}
\end{widetext}
Following the definition of the connected correlation function we arrive at 
\begin{widetext}
\begin{eqnarray}
\label{TA3}
M(t) &=& \beta D(t) + C_1(t) + C_2(t) + C_3(t) \nonumber\\
&+& \frac{\beta}{2t} \langle x_t-x_0 \rangle\,\langle x_t-x_0 \rangle
+ \frac{\beta}{2t}\int_0 ^{t} ds\,\int \id r \,\Gamma(r-s)\,\langle
x_t-x_0 \rangle\,\langle\dot{v_r}\rangle\nonumber\\
&+& \frac{\beta}{4t}\int_0 ^{t} ds \int \id r \int \id
u \Gamma(r-s)\{\gamma(r-u)-\gamma(u-r)\}\,\langle
x_t-x_0 \rangle\,\langle v_u \rangle \nonumber\\
&-&\frac{\beta}{2t}\int_0 ^{t} \int \id r \,\Gamma(r-s) \, \langle
F_r(x_r)\rangle\,\langle x_t-x_0 \rangle\;ds\nonumber\\
\end{eqnarray}
\end{widetext}
From the Langevin equation we have 
\begin{equation}
\langle \dot{v_r} \rangle = -\int du \gamma(r-u)\, \langle v_u \rangle + \langle F_r(x_r) \rangle \nonumber\\
\end{equation}
After substituting this in \eqref{TA3}, all the terms cancel out each other and only the first line will remain. The cancellation for the 
forcing term is clear; and the other terms follow as:
\begin{widetext}
\begin{eqnarray}
-\frac{\beta}{2}\int_0 ^{t} ds \int \id r \int \id u \Gamma(r-s)\,\gamma(r-u)\,\langle v_u \rangle& =& 
-\frac{\beta}{4}\int_0 ^{t} ds \int \id r \int \id u \Gamma(r-s)\,\gamma(r-u)\,\langle v_u \rangle \nonumber\\
&-&\frac{\beta}{4}\int_0 ^{t} ds \int \id r \int \id u \Gamma(r-s)\,\gamma(r-u)\,\langle v_u \rangle \nonumber\\
&-&\frac{\beta}{4}\int_0 ^{t} ds \int \id r \int \id u \Gamma(r-s)\,\gamma(u-r)\,\langle v_u \rangle \nonumber\\
&+&\frac{\beta}{4}\int_0 ^{t} ds \int \id r \int \id u \Gamma(r-s)\,\gamma(u-r)\,\langle v_u \rangle \nonumber\\
&=& - \frac{\beta}{4t}\int_0 ^{t} ds \int \id r \int \id u \Gamma(r-s)\{\gamma(r-u)+\gamma(u-r)\}\,\langle v_u \rangle \nonumber\\
&-& \frac{\beta}{4t}\int_0 ^{t} ds \int \id r \int \id u \Gamma(r-s)\{\gamma(r-u)-\gamma(u-r)\}\,\langle v_u \rangle \nonumber\\
&=& - \frac{\beta}{2} \langle x_t-x_0 \rangle - \frac{\beta}{4t}\int_0 ^{t} ds \int \id r \int \id u \Gamma(r-s)\{\gamma(r-u)-\gamma(u-r)\}\,\langle v_u \rangle \nonumber\\
\end{eqnarray}
\end{widetext}

\section{Simulation of colored Gaussian noise}\label{sim}

We sketch the algorithm to numerically generate a stationary Gaussian
coloured noise $\xi_t$ for a given time-correlation function $\gamma$.
The strategy is to transform to Fourier space as e.g in \cite{cng,cng12}, then to
simulate  it there as e.g. in \cite{cngb}, and then to transform the solution back to
real space.  We next explain that scheme in more detail.\\ Suppose the noise satisfies
\begin{eqnarray}
\label{expect}
\left< \xi(t) \right> & = & 0 \nonumber\\
\left< \xi(t)\xi(t') \right> & = & \gamma(|t-t'|)
 \end{eqnarray}
In discrete Fourier space the noise can be constructed as
 \begin{equation}
\label{Inv}
\tilde{\xi}(\omega_\mu) = \sqrt{ N \tilde{\gamma}(\omega_\mu)} \;\alpha_\mu \;\;\; ,\;\  \mu=0,...,N-1
\end{equation}
with $N$ even, and $\tilde{\xi}(\omega_\mu)$ and $\tilde{\gamma}(\omega_\mu)$ being the Fourier transforms of $\xi(t)$ and $\gamma(t)$,
respectively; the $\omega_\mu$ is defined as
\begin{equation}
\omega_\mu=2\pi\frac{\mu-\frac{N}{2}}{N\Delta},\quad \omega_{N-\mu} = -\omega_{\mu}
\end{equation}
where $\Delta$ is the sampling interval of time.  Finally,
 $\alpha_\mu$ is a Gaussian complex random number in Fourier space
 with zero mean and correlation, \cite{cng2},
\begin{equation}
\label{avali}
 \left< \alpha_\mu\alpha_\nu\right>=\delta_{\mu,N-\nu},\quad \alpha^*_\mu=\alpha_{N-\mu}
\end{equation}
To generate a Gaussian complex random number with correlation given in
{\ref{avali}), we write $\alpha_\mu = a_\mu+ib_\mu$ in terms of its
real and imaginary parts: $\alpha_\mu=a_\mu+ib_\mu$ if $\mu>N/2$ and
otherwise $\alpha_\mu=a_{N-\mu}-ib_{N-\mu}$.  Here, $a$ and $b$ are
two Gaussian real random numbers which are uncorrelated and have zero
mean and covariance
\begin{eqnarray}
 \left<a_\mu a_\nu\right> = \frac{1}{2}\;(\delta_{\mu,\nu}
 + \delta_{\mu,N-\nu})\nonumber\\
\left<b_\mu b_\nu\right> = \frac{1}{2}\;(\delta_{\mu,\nu} - \delta_{\mu,N-\nu})\nonumber\\
\end{eqnarray}
It is then straightforward to do the inverse Fourier transform
\[
 \xi(t_k)  =  \frac{1}{N} \sum_{\mu=0}^{N-1}\tilde{\xi}(\omega_\mu) e^{i\omega_\mu t_k}
\]
and to see that the correlations reproduce (\ref{expect}).



\begin{thebibliography}{10}

\bibitem{proc}
M.~Baiesi, C.~Maes and B.~Wynants, The modified Sutherland-Einstein relation for diffusive non-equilibria. Proc. Royal Soc. A {\bf 467}, 2792--2809 (2011).

\bibitem{up}
M.~Baiesi and  C.~Maes, An update on nonequilibrium linear response.
{\tt arXiv:1205.4157v2 [cond-mat.stat-mech]} to appear in New Journal of Physics.

\bibitem{fdr}
M.~Baiesi, E.~Boksenbojm, C.~Maes and B.~Wynants, Nonequilibrium Linear Response for Markov Dynamics,II: Inertial Dynamics. J. Stat.Phys. {\bf 139}, 492--505 (2010).

\bibitem{har}
T.~Harada and S.-Y.~Sasa, Equality connecting energy dissipation with
  violation of fluctuation-response relation. Phys. Rev. Lett. {\bf 95}, 130602 (2005).

\bibitem{villamaina}
D. Villamaina, A. Baldassarri, A. Puglisi, A. Vulpiani, J. Stat. Mech. P07024 (2009)

\bibitem{crisanti}
A. Crisanti, A. Puglisi, and D. Villamaina, Phys. Rev. E {\bf 85}, 061127 (2012)

\bibitem{pro}
P.~Bohec, F.~Gallet, C.~Maes, S.~Safaverdi, P.~Visco and F.Van Wijland, Probing active forces via a fluctuation-dissipation relation.  {\tt arXiv:1203.3571v1 [cond-mat.soft]}.

\bibitem{zwa61}
R.~Zwanzig, Memory effects in irreversible thermodynamics, Phys. Rev. {\bf 124}, 983 (1961).

  \bibitem{mor65}
H.~Mori, Transport, collective motion, and Brownian motion. Prog. Theor. Phys. {\bf 33}, 423--455 (1965).

\bibitem{pan}
D.~Panja, Generalized Langevin equation
formulation for anomalous polymer
dynamics. J. Stat. Mech.  L02001 (2010).

\bibitem{mae03}
C.~Maes and K.~Neto\v{c}n\'{y}, Time-reversal and entropy. J. Stat. Phys. {\bf 110}, 269--310 (2003).

\bibitem{kub66}
R.~Kubo, The fluctuation-dissipation theorem. Rep. Prog. Phys. {\bf 29}, 255--284 (1966).

\bibitem{tod92}
R.~Kubo, M.~Toda and N.~Hashitsume, {\em Statistical Physics: Nonequilibrium
  statistical mechanics} 2nd ed vol~2 (Springer), 1992.

 \bibitem{brad}
J.M.~Brader, M.~Siebenbuerger, M.~Ballauff, K.~Reinheimer, M.~Wilhelm, S.J.~Frey, F.~Weysser and M.~Fuchs, Nonlinear response of dense colloidal suspensions under oscillatory shear:
Mode-coupling theory and FT-rheology experiments. Phys. Rev. E {\bf 82}, 061401 (2010)

\bibitem{krug1}
M.~Kr\"uger and M.~Fuchs, Nonequilibrium fluctuation-dissipation relations of interacting Brownian particles driven by shear. Phys. Rev. E {\bf 81}, 011408 (2010).

\bibitem{krug2}
M.~Kr\"uger and M.~Fuchs, Non-Equilibrium relation between mobility and diffusivity of interacting Brownian particles under shear. Prog. Theor. Phys. Suppl. {\bf 184}, 172-–186 (2010).

\bibitem{ber}
L.~Berthier and J.L.~Barrat, Nonequilibrium dynamics and fluctuation-dissipation relation in a sheared fluid. J. Chem. Phys. {\bf 116}, 6228 (2002).

\bibitem{camillep}
C.~Aron, G.~Biroli and L.~F.~Cugliandolo, Symmetries of generating functionals of Langevin processes with colored multiplicative noise. J. Stat. Mech. P11018 (2010).
\bibitem{camillet}
C.~Aron, Classical and quantum out-of-equilibrium dynamics. Formalism and applications. PhD thesis, Universit\'e Pierre et Marie Curie (2010).
\bibitem{deu}
J.M.~Deutsch and Onuttom Narayan, Energy dissipation and fluctuation response for particles in fluids.
Phys. Rev. E {\bf 74}, 026112 (2006).
\bibitem{fox}
R.F.~Fox, The generalized Langevin equation with Gaussian fluctuations. J. Math. Phys. {\bf 18}, 2331--2335 (1977).

\bibitem{han1}
P.~H\"anggi, Correlation Functions and Masterequations of Generalized (Non-Markovian) Langevin Equations. Z. Phys. B {\bf 31}, 407--416 (1978).
\bibitem{han}
P.~H\"anggi, Path integral solutions for non-Markovian processes. Z. Phys. B --- Condensed Matter {\bf 75}, 275--281 (1989).
\bibitem{nov}
E.A.~Novikov, Zh. Eksp. Teor. Fiz. {\bf 47}, 1919 (1964); [Sov.
Phys. JETP {\bf 20}, 1290 (1965)].

\bibitem{fur}
K.~Furutsu, J. Res. Natl. Bur. Stand., Sect. D {\bf 67}, 303 (1963).

\bibitem{don}
M.~D.~Donsker, On function space integrals. In
{\it Analysis in Function Space}, Eds. W.T.~Martin and I.~Segal, MIT Press; pp 17–-30 (1964).


\bibitem{dyk}
M.~I.~Dykman and I.~B.~Schwartz,
 Large rare fluctuations in systems with delayed dissipation. Phys. Rev. E {\bf 86}, 031145 (2012).

\bibitem{ckp}
L.F.~Cugliandolo, J.~Kurchan and G.~Parisi,
Off equilibrium dynamics and ageing in unfrustrated systems. J. Physique I (France) {\bf 4}, 1641 (1994).

\bibitem{noe}
N.~Pottier,
Relaxation time distributions for an anomalously diffusing particle. Physica A {\bf  390}, 2863 (2011).


\bibitem{cng}
Kun L\"u and Jing-Dong Bao, Numerical simulation of generalized Langevin equation with arbitrary correlated noise. Phys.Rev. E {\bf 72}, 067701 (2005).

\bibitem{cng12}
J.~García-Ojalvo, J.M.~Sancho, Noise in Spatially Extended Systems. {\it Springer-Verlag, New York} (1999).
\bibitem{cngb}
W.H.~Press,  W.T.~Vetterling, B.P.~Flannery and S.A.~Teukolsky, Numerical Recipes: The Art of Scientific Computing {\it Cambridge University Press}, 2007.


\bibitem{cng2}
J.M.~ Porr\`a, Ke-Gang Wang, and Jaume Masoliver, Generalized Langevin equations: Anomalous diffusion and probability distributions. Phys. Rev. E {\bf 53}, 5872–-5881 (1996).








 






\end{thebibliography}
\end{document}